\documentclass[10pt, conference]{IEEEtran}
\usepackage{adjustbox}
\usepackage[normalem]{ulem}
\useunder{\uline}{\ul}{}

\usepackage{cite}

\usepackage[printwatermark]{xwatermark}
\usepackage{xcolor}
\usepackage{graphicx}
\usepackage{lipsum}

\newwatermark[allpages,color=gray!50,angle=45,scale=3,xpos=0,ypos=0]{DRAFT}

\ifCLASSINFOpdf
\else
\fi

\usepackage{graphicx}
\usepackage{multirow}
\usepackage{stfloats}

\usepackage{pdflscape}
\usepackage{array,booktabs}
\usepackage[none]{hyphenat}

\begin{document}

\newcommand{\ra}[1]{\renewcommand{\arraystretch}{#1}}
\newcolumntype{L}{>{\raggedright\arraybackslash}p}
\newcolumntype{C}{>{\centering\arraybackslash}p}
\newcolumntype{F}{>{\arraybackslash}p}
\newcolumntype{R}{>{\raggedleft\arraybackslash}p}

\title{Sustainability: Delivering Agility's Promise}

\author{\IEEEauthorblockN{Jutta Eckstein}
\IEEEauthorblockA{
Email: jutta@jeckstein.com}
\and
\IEEEauthorblockN{Claudia de O. Melo}
\IEEEauthorblockA{
Email: claudia.melo.prof@gmail.com}
}

\maketitle

\begin{abstract}
    Sustainability is a promise by agile development, as it is part of both the Agile Alliance's and the Scrum Alliance's vision. Thus far, not much has been delivered on this promise. This paper explores the Agile Manifesto and points out how agility could contribute to sustainability in its three dimensions – social, economic, and environmental. Additionally, this paper provides some sample cases of companies focusing on both sustainability (partially or holistically) and agile development.
\end{abstract}

\section{Introduction}
\label{sec:introduction}
The two major agile organisations, the Agile Alliance and the Scrum Alliance, both promise in their vision statements that sustainability is one of their core goals:
\begin{verse}
Agile Alliance is a nonprofit organisation committed to supporting people who explore and apply Agile values, principles, and practices to make building software solutions more effective, humane, and sustainable \cite{1}.\\
Scrum Alliance\textregistered is a nonprofit organisation that is guiding and inspiring individuals, leaders, and organisations with agile practices, principles, and values to help create workplaces that are joyful, prosperous, and sustainable \cite{2}.
\end{verse}
 
If we want to support building software solutions to be more effective, humane, and sustainable or to help create workplaces that are joyful, prosperous, and sustainable we have to aim (among other things) for sustainability. However, thus far not much has been done for approaching this aim.

In this paper, we are going to provide a new lens in order to understand the Agile Manifesto under the premise the agile approach wants to fulfil its promise for sustainability and we will provide various case studies of companies attempting to use agile development to contribute to sustainability.

The paper is structured as follows: we will at first examine the various definitions of sustainability and explore both how the business and Information and Communication Technologies (ICT) approach and classify sustainability. We will then take a close look at the principles defined by the Agile Manifesto in order to find out how they can support sustainable development \cite{12}. Next, we present various case studies of companies either addressing sustainability partially or holistically by leveraging it with an agile approach. In the conclusion, we will take a critical look at sustainability initiatives before we will provide an outlook on the (hopefully) not so far future.

\section{Sustainability}
\label{sec:sustainability}
There are several definitions for sustainability and nuances across the spectrum of sustainable use, sustainable development and sustainability \cite{Hilty2015}. The most famous and frequently definition adopted to frame discussions around sustainability is provided by the Brundtland report \cite{21}:
\begin{quote}
   ``Sustainable development is development that meets the needs of the present without compromising the ability of future generations to meet their own needs. It contains within it two key concepts: the concept of 'needs', in particular the essential needs of the world's poor, to which overriding priority should be given; and the idea of limitations imposed by the state of technology and social organisation on the environment's ability to meet present and future needs''.
\end{quote}

The Brundtland report suggests how economic and social development should be defined and calls all countries to action:
\begin{quote}
Thus the goals of economic and social development must be defined in terms of sustainability in all countries - developed or developing, market-oriented or centrally planned. Interpretations will vary, but must share certain general features and must flow from a consensus on the basic concept of sustainable development and on a broad strategic framework for achieving it.
\end{quote}

Finally, the report describes how a path towards sustainability should look like, bringing the concept of physical sustainability (related to living under the laws of nature and minimising the impact on the physical environment) and its connection to intra- and inter- generational social equity:
\begin{quotation}
    ``Development involves a progressive transformation of economy and society. A development path that is sustainable in a physical sense could theoretically be pursued even in a rigid social and political setting. But physical sustainability cannot be secured unless development policies pay attention to such considerations as changes in access to resources and in the distribution of costs and benefits. Even the narrow notion of physical sustainability implies a concern for social equity between generations, a concern that must logically be extended to equity within each generation''.
\end{quotation}

According to the these definitions, sustainability is about taking long-term responsibility for your action and reaches further than energy consumption and pollution as it is often casually understood. 

Other important concepts that seek to explain sustainability are: the three-pillar model and the triple bottom line. The \textbf{three-pillar model} depicts sustainability by synthesising social, economic, and environmental concerns. It is the model most widely used, for example, it is the definition given by Wikipedia and also used on the 2005 World Summit on Social Development \cite{3,4}. However, as explained in \cite{Purvis2019}, ``the conceptual foundations of this model are far from clear and there appears to be no singular source from which it derives''.

The \textbf{triple bottom line} is an accounting framework that seeks to broadening the notion of a company bottom line by introducing a full cost accounting. A single bottom line is the company's profit (if negative, loss) in an accounting period. A triple bottom line adds social and environmental (ecological) concerns to the accounting. If a corporation has a monetary profit, but it causes thousands of deaths or pollutes a river, and the government ends up spending taxpayer money on health care and river clean-up, the triple bottom line needs to account for these cost-benefit analysis too.  

The triple bottom line is also known by the phrase \textbf{``people, planet, and profit''} and was coined by John Elkington in 1994 while at SustainAbility (a British consultancy). A triple bottom line company seeks to gauge a corporation's level of commitment to corporate social responsibility and its impact on the environment over time \cite{23}. 

Another important framework that also articulates sustainability is the United Nations 2030 Agenda and the 17 Sustainable Development Goals: ``The Sustainable Development Goals are a universal call to action to end poverty, protect the planet and improve the lives and prospects of everyone, everywhere'' \cite{6}.

\begin{figure}[h]
    \centering
    \includegraphics[width=0.5\textwidth]{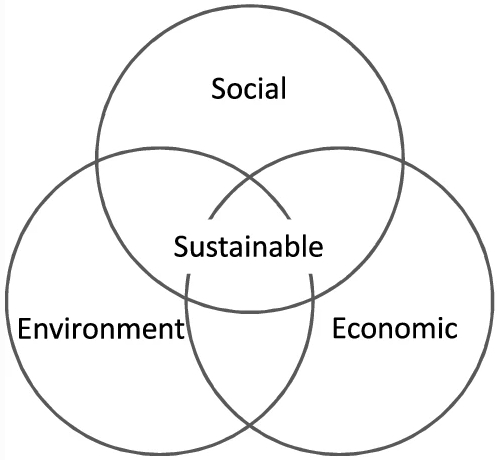}
    \caption{The three dimensions of sustainability \cite{Purvis2019}}
    \label{fig:3pillars}
\end{figure}

Also these sustainable development goals are founded in the three pillar model: social (the aim of ending poverty), environmental (protecting the planet), and economic (improving the lives and prospects of everyone, everywhere). Therefore, throughout this article we will use the three-pillar model as the definition for sustainability, as illustrated by Figure \ref{fig:3pillars}.

Even with the definition of the three pillars, it has always to be understood that sustainability is highly interconnected that any elaboration on sustainability requires a holistic perspective because all actions are interdependent \cite{Borowski2018}:
\begin{quote}
    ``All definitions of sustainable development require that we see the world as a system—a system that connects space; and a system that connects time. When you think of the world as a system over space, you grow to understand that air pollution from North America affects air quality in Asia, and that pesticides sprayed in Argentina could harm fish stocks off the coast of Australia. And when you think of the world as a system over time, you start to realise that the decisions our grandparents made about how to farm the land continue to affect agricultural practice today; and the economic policies we endorse today will have an impact on urban poverty when our children are adults''.
\end{quote}

\subsection{Business and Sustainability}
According to the dictionary \cite{24}, ``Business pertains broadly to commercial, financial, and industrial activity, and more narrowly to specific fields or firms engaging in this activity''.
    
However, also the business and its self-conception is changing. In particular, this became visible, by the Business Roundtable, an association of chief executive officers of leading companies in the USA, refining its statement on the purpose of a corporation in 2019. This refined statement made a shift from the focus on satisfying shareholders by making money to a more holistic understanding and aiming for satisfying customers, employees, communities, suppliers, and shareholders equally. This shift comprehends an understanding that financial success is not the sole purpose of business. 

For example, the statement says \cite{25}: 
\begin{quotation}
    ``...when it comes to addressing difficult economic, environmental and societal challenges, these companies are starting in their own backyards – partnering with communities to provide the investment and innovative solutions needed to revitalize local economies and improve lives. 
    
    These investments and initiatives aren't just about doing good; they're about doing good business and creating a thriving economy with greater opportunity for all.'' 
\end{quotation}

Although, when talking about sustainability, the Business Roundtable refers only to energy and environment, they show an understanding of sustainability as it is defined by the Brundtland report \cite{26, 21} by pointing out the importance and interdependence of the economic, environmental, and societal challenges. 


At the organisational level, there are concrete movements aiming to answer the question of how an organisation can be more balanced across the three sustainability pillars. They recognise that capitalism, as it's currently practised, is starting to run into fundamental structural problems. 

In the US, \textbf{B Lab} \cite{BCorp} has worked to create a certification of ``social and environmental performance'' to evaluate for-profit companies. B Lab certification requires companies to meet social sustainability and environmental performance, and accountability standards, as well as being transparent to the public according to the score they receive on the assessment. Later on, they also developed the concept of ``benefit corporations'', companies that legally commit themselves to honour moral values, while pursuing the standard capitalist goal of maximising profits. A few examples of well-known benefit corporations include Method, Kickstarter, Plum Organics, King Arthur Flour, Patagonia, Solberg Manufacturing, Laureate Education, and Altschool.

In Europe, \textbf{Economy for the Common Good (ECG)} \cite{ECG} has similar goals. ECG is an economic model, which makes the Common Good, a good life for everyone on a healthy planet, its primary goal and purpose. The Economy for the Common Good (ECG) was initiated in 2010 by economic reformist and author Christian Felber, together with a group of Austrian pioneer enterprises. ECG is an ethical model for society and the economy with the goal of reorienting the free market economy through a democratic process towards common good values. According to Felber, it seeks to address a capitalist system that ``creates a number of serious problems: unemployment, inequality, poverty, exclusion, hunger, environmental degradation and climate change''. The solution is an economic system that ``places human beings and all living entities at the centre of economic activity''.

At the heart of ECG lies the idea that values-driven businesses are mindful of and committed to: 1) Human Dignity; 2) Solidarity and Social Justice; 3) Environmental Sustainability, and 4) Transparency and Co-Determination. ECG is currently supported by more than 1800 enterprises in 40 countries such as Sparda-Bank Munich, VAUDE, Sonnentor, and taz (German newspaper), about 250 have created a Common Good Balance Sheet. This balance sheet is a scorecard that measures companies based on their preservation of those 4 fundamental values, considering 5 key stakeholders: Suppliers; Owners/Equity/Financial Service Providers; Employees/Co-worker employers; Customers/Other Companies; and Social Environment. 

\subsection{ICT/Technology and Sustainability}
\label{sec:csandsustainability}
The idea of using Information and Communication Technologies (ICT) to address sustainability issues has been investigated in a number of interdisciplinary fields that combine ICT with Environmental and/or Social sciences. Amongst these are Environmental informatics, Computational sustainability, Sustainable Human-Computer Interaction, Green IT/ICT, or ICT for Sustainability \cite{Hilty2015}.

The contributions of these fields are manifold: monitoring the environment; understanding complex systems; data-sharing, and consensus-building; decision support for the management of natural resources; reducing the environmental impact of ICT hardware and software; enabling sustainable patterns of production and consumption; and understanding and using ICT as a transformational technology.

For these different areas, there is a common understanding that ICT can be used to reduce its own footprint and to support sustainable patterns \cite{Hilty2015}:
\begin{itemize}
    \item \textbf{Sustainability in ICT} (also referred as Green in ICT): Making ICT goods and services more sustainable over their whole life cycle, mainly by reducing the energy and material flows they invoke.
    \item \textbf{Sustainability by ICT} (also referred as Green by ICT): Creating, enabling, and encouraging sustainable patterns of production and consumption
\end{itemize}

The problem with this differentiation is that often sustainability in and by ICT are intertwined. When analysing the impact of certain technology, looking for a final positive or negative conclusion, it is not possible to assess it in isolated contexts to make a decision. A positive positive occurs when the effect of a sustainable activity on the social fabric of the community causes well-being of the individuals and families considering the three pillars \cite{deSousa2019}. Thus, assessing the ICT impact needs to consider that  \cite{Hilty2015}:
\begin{quote}
``Sustainable development [...] is defined on a global level, which implies that any analysis or assessment must ultimately take a macro-level perspective. Isolated actions cannot be considered part of the problem, nor part of the potential solution, unless there is a procedure in place for systematically assessing the macrolevel impacts''.
\end{quote}

Therefore, the assessment of specific technology impact on sustainable development must consider its nuances and inter-dependencies on multiple levels, that might be on both spaces of ``in'' and ``by''. A good example is provided in \cite{ICT4S2013}, p 284:
\begin{quote}
    ``[...] history of technology has shown that increased energy efficiency does not automatically contribute to sustainable development. Only with targeted efforts on the part of politics, industry and consumers will it be possible to unleash the true potential of ICT to create a more sustainable society''.
\end{quote}

\section{Agile and Sustainability}
\label{sec:agileandsustainability}
If agile is really aiming for sustainability, as it is suggested by the vision of both the Agile Alliance and the Scrum Alliance, then we should take a look at the principles of the Agile Manifesto to understand how these principles can contribute to or guide sustainability \cite{12}. Sustainability requires a much broader view to integrate the environmental, economic, and social perspectives. 

\subsection{Our highest priority is to satisfy the customer through early and continuous delivery of valuable software.}
At the core of this very first principle is continuous learning by \textit{focusing constantly on the customer}. Only through continuous delivery you will be able to keep adjusting the system to the customer's satisfaction. Broadening the perspective, and looking at this principle through a sustainability lens, that is taking also the environmental and social aspect into account, shifts also the meaning of ``valuable'' software. The value is not only defined by the economic benefit for the customer but also by the social and environmental improvements. 

Moreover, by broadening the perspective, we'll find that it will become even harder to come up with a perfect solution right away. For example, measuring the carbon footprint of the system you're building will teach you what aspects you need to adapt, rethink, and redo. Thus, adding two more dimensions (environmental and social) increases the complexity of development and thus, \textit{continuous learning} is even more important to uncover emergent practices \cite{Kurtz2003}.

\subsection{Welcome changing requirements, even late in development. Agile processes harness change for the customer's competitive advantage.}
\begin{figure}[h]
    \centering
    \includegraphics[width=0.4\textwidth]{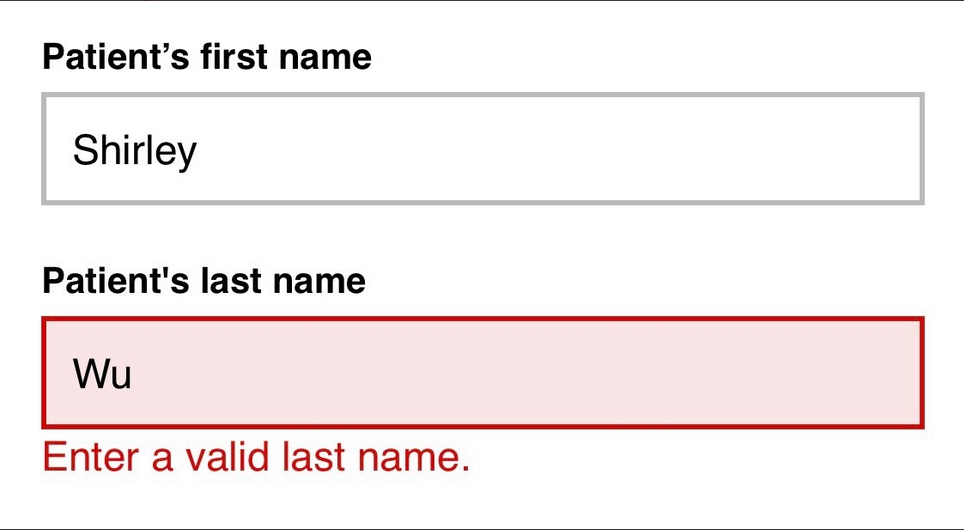}
    \caption{Application requiring two characters for last name}
    \label{fig:2charsLastname}
\end{figure}

This second principle addresses sustainability by ICT, by \textit{focusing on the customer} (like the principle before). At first sight, the ``customer's competitive advantage'' sounds like it can only aim for economical success. Yet, the advantage can also be built by a reputation of the product as being environmental and social friendly. As a counter example, the authors have seen websites that require the specification of the first and last name of the user (Figure \ref{fig:2charsLastname}\footnote{\url{https://twitter.com/shirleyywu/status/1300628412466298881?s=20}}). However, both need to be at least three characters long. If the developers would have been socially aware they would have understood that there are many people (especially in Asia) whose names are only two characters long. This is only a minor example but still shows how broadening the view can make a difference for the customer - by gaining a good or bad reputation. 

To provide the customer this kind of competitive advantage, we have to answer:
\begin{itemize}
    \item How is the product that we're creating helping the world?
    \item How is it part of a wider sustainable solution?
    \item How can we ensure the product is inclusive?
    \item How can the product even improve the environment?
    \item How can we ensure the product itself is environment friendly?
\end{itemize} 

\subsection{Deliver working software frequently, from a couple of weeks to a couple of months, with a preference to the shorter timescale.}
Similar to the first principle, also this one focuses on \textit{continuous learning}. The difference is that this third principle asks to establish a regular cadence for feedback. The shorter the feedback cycle, the better, because then we can keep the focus on the learning about the changes in the system by the last delivery. This principle is important for tackling sustainability because the knowledge, wisdom, and experience is steadily increasing in all three dimensions (economical, social, and environmental) and this learning has to be reflected in the system. 

For example, although we might have designed the system to be highly accessible, only by getting the feedback from the people who are in the need of the accessibility will we know how well the system performs. Similarly, we can learn from the real carbon footprint of the system only when we know how it performs in reality. 

To continuously learn from the delivery you should regularly ask yourself questions as \cite{Frick2016}:
\begin{itemize}
    \item Does the system work for people with disabilities?
    \item Does the system work for people using older devices?
    \item Does the system provide the best possible performance with the least amount of resources across devices and platforms?
\end{itemize}

\subsection{Business people and developers must work together daily throughout the project.}
This is the fourth principle of the Agile Manifesto and points out the importance of collaboration between the ones creating the product or service (developers) and the ones who want to offer that product or service to the market. Continuous collaboration enables \textit{self-organisation} around the system created and this allows to discover and actually meet the market needs. 

The collaboration between business people and developers is not market focused only (anymore), but keeps the wider perspective of addressing all three dimensions of sustainability. This includes a reflection on the market we serve, disrupt, and/or create. For example, in the last decades car manufacturers did create a market for heavy and environmental unfriendly cars (the sport Utility Vehicles, SUVs) and hesitated on the other hand to create a market for electric cars with the argument that this market does not exist \cite{MORTSIEFER2017}.

Thus, this principle is a continuous reminder to consider all aspects of sustainability in our daily work.

\subsection{Build projects around motivated individuals. Give them the environment and support they need, and trust them to get the job done.}
This is the principle number five and directly mentions the environment. First of all, this principle is a call to action for sustainability in ICT. At first sight, the social pillar is emphasising the importance of supporting the individuals for example, by paying them fair (also despite any difference in i.e., race, gender, religion, location, etc.). Providing an environment that helps the individuals to \textit{self-organise} for getting their job done includes the technical tools yet, also the safety of the environment: the individuals are not put at risk by e.g. a polluted or toxic work-space (think of asbestos, colours, or carpets that evaporate toxic emissions). 

Moreover, the environment can also contribute, neutralise, or reduce the carbon footprint. For example, by the energy that is used, the availability of natural light, the existence of natural plants, or the commute that is required. Finally, from an economic point of view, this principle also requests that the environment allows every individual in the same way to prosper and grow by offering life-long learning.

\subsection{The most efficient and effective method of conveying information to and within a development team is face-to-face conversation.}
This sixth principle puts direct conversation at the core. Face-to-face conversation provides \textit{transparency} particularly from the social and economic point of view. If the system you are building should support for example, (individual) growth and development or / and inclusiveness, then you will get qualitatively better feedback from the users if talking to them directly. Direct observation, eye contact, mimic, and gesture provide additional information how well the system is serving the users.

This principle invites as well to build a community with everyone involved in making the product sustainable. The community reflects in the process and in the product it is an interdependent relationship.

\subsection{Working software is the primary measure of progress.}
This principle aims for \textit{transparency}. Certainly, it is important to think ideas through however, you will only know if they keep their promises when hitting reality. This includes to regularly examine the system for unintended consequences (also known as the precautionary principle \cite{Riordan2013}). Therefore, it is also the responsibility of an agile team to prevent and monitor for any unintended consequences in order to address them \cite{Tenner1997}. 

Therefore, never underestimate the importance of feedback for example, on energy consumption on the running system, theory and reality do not always show the same results. From the social perspective, regularly examine the working system guided by questions as:
\begin{itemize}
    \item Does the system work in the same way for everyone independent if i.e., the gender differs or it is used by people from different races or ethnicity?
    \item Would a domestic abuser find possibilities to do harm to the system but more so to other people?
    \item Can the system be used by a government to oppress?
\end{itemize}
We suggest to take a look at other questions, for example the ones offered by the Ethical Explorer toolkit \cite{EthicalExplorer}.

\subsection{Agile processes promote sustainable development. The sponsors, developers, and users should be able to maintain a constant pace indefinitely.}
This is the eighths of twelve principles and the only one that calls for sustainable development explicitly. It focuses foremost on sustainability in ICT. Thus far, this principle has mainly been understood from the social aspect for example, calls for the '40 hour week' have been argued with the reference to this principle \cite{Beck2000}.

Using the lens of sustainability, it asks that developers \textit{learn continuously} about all three pillars and take the learning into account: environment, social, and economic. Therefore, having a constant pace in mind, developing the product or service should not lead to burnout of the developers (social), to reduction of resources (environment), or to overspending financially (economic). For example, the features developed should take the diversity and the abilities of the users into account, the energy used creating the product should be renewable, and the so-called feature-creep should be avoided.

Feature-creep, the inclusion of unnecessary features and the development of features that do not adhere to the intended design and architecture leads to inefficiency of the overall system. Moreover, the utilised capacity of the CPU in idle mode or the demand for storage are reasons for software getting slower. 

Therefore, sustainable development has to be taken into account by the developers and the sponsors to ensure that, for example, the user is not required to invest in a new hardware regularly. Modularity and allowing the user to determine which modules to buy, install, and use take into account that not every user will need every possible innovation (at least not on every device).

\subsection{Continuous attention to technical excellence and good design enhances agility.}

Developments in sustainability are progressing fast - from new understandings of algorithm bias (social aspect), or of where most energy is used and how it could be reduced i.e. by storing data locally and reducing network traffic (environmental aspect) to better ways for finding out which features are actually used and so - to ensure that money is not wasted by feature-creep.

Thus, this ninth principle, is a call for \textit{continuous learning} and for keeping technologically up-to-date and taking into account new learning and development that enable good design: now good in the sense of sustainability. This ensures the principle supports sustainability in ICT.

\subsection{Simplicity--the art of maximising the amount of work not done--is essential.}

The tenth principle often requires a second read before it is understood. In general, it addresses the feature-creep mentioned above. So, when developing software, we need to pay attention to what is really needed in the product. Looking through a sustainability lens at this principle, we need to examine for example, how does this new feature fit in the system without increasing the energy consumption? Using Scrum as an example, sustainability or rather energy consumption needs to be considered during backlog refinement, sprint planning, by the definition of done, as well as monitored through tests.

The German Federal Environment Office created and published in 2020 a label (or certificate) ``Blauer Engel'' for resources and energy-efficient software stand-alone products, based on a criteria catalogue jointly developed by the university of Trier and Zurich \cite{13,14}. The label focuses on energy-efficiency, conservative resource consumption, and transparent interfaces. The plan is to develop a similar label for cloud-based software. However, already today, the criteria catalogue can guide a team to develop a sustainable system. For example, the following questions should be regularly discussed \cite{14}:
\begin{itemize}
    \item How much electricity does the hardware consume when the software product is used to execute a standard usage scenario?
    \item Does the software product use only those hardware capacities required for running the functions demanded by the individual user? Does the software product provide sufficient support when users adapt it to their needs?
    \item Can the software product (including all programs, data, and documentation including manuals) be purchased, installed, and operated without transporting physical storage media (including paper) or other materials goods (including packaging)?
    \item To what extent does the software product contribute to efficient management of the resources it uses during operation?
\end{itemize}
Thus, this principle requires \textit{transparency} for features that are needed (and used) and those that are not.

\subsection{The best architectures, requirements, and designs emerge from self-organising teams.}
The eleventh principle points out the benefits of \textit{self-organising} teams. This includes that every team member is invited to speak up and make their contribution to the architecture, requirements, and design -  independent of other characteristics of that team member (social perspective). Similarly, all team members will get the same fair chance to progress on the their career by getting equal support through training, mentoring, or coaching (economic perspective).

\subsection{At regular intervals, the team reflects on how to become more effective, then tunes and adjusts its behaviour accordingly.}
The twelfth and final principle asks teams to run regular retrospectives. Both the reflection and behavioural adjustments should take (also) sustainability aspects into account: how can the team tune and adjust its behaviour to become more effective regarding the three pillars - environmental, social, and economical? Dedicating time regularly for reflecting on sustainability will lead to continuous improvements. This last principle combines the quest for teams to \textit{self-organise} in order to \textit{learn continuously} by making their effectiveness with the \textit{focus on the customer} \textit{transparent}.

\subsection{Summary of the Agile Manifesto's Perspective on Sustainability}
Implementing sustainable development goals requires approaching wicked problems, i.e. complex, non-linear, dynamic challenges in situations of insufficient resources, incomplete information, emerging risks and threats, and fast changing environments \cite{Melo2019}. Examining the principles of the Agile Manifesto shows how an agile approach indeed promotes (or can promote) sustainable development. It might be surprising how much guidance the principles can provide although they have been defined originally with the focus on software development only. 

Concentrating continuously on inspect and adapt allows sustainable systems to emerge. An agile, cross-functional team integrates different perspectives on the emerging system and has this way the possibility to design solutions for sustainability. The disciplined approach provided by agile development enables the team to permanently learn from their delivery, to measure the outcome according its environmental, social, and economic impact and take according actions for adjustments. 

However, taking all three dimensions of sustainability into account leads to higher complexity, so an agile approach also comes in handy for addressing this complexity by breaking down the problems and using an inspect and adapt approach for making them simpler. 

The role of Agility is not to save the world, but to provide a value system based on \textit{transparency, constant customer focus, self-organisation, and continuous learning} that leads into sustainable thinking and offers an approach that supports putting this sustainable thinking into action.

\section{Case Studies: Leveraging Agility for Sustainability}
\label{sec:casestudies}
Although the Agile Manifesto \cite{12} originates in 2001 and the Brundtland report \cite{21} in 1987, the combination of agility and sustainability is just in its beginnings. There are some sample companies being conscious of one of the three pillars and even fewer sample companies taking a holistic approach on combining sustainability and agility. In this section we will provide some sample case studies for both attempts - agility and one of the three pillars as well as company-wide agility and sustainability using a holistic approach.

We want to point out that for the following case studies, as for other examples, there is no such thing as a ``perfect'' company, neither in terms of sustainability nor in terms of agility. However, these case studies still can serve as examples of possible steps to take in order to leverage agility for sustainability.

\subsection{Agility and Partial Sustainability}
\label{sec:impact}
In this section, we explore different examples from companies applying agility on one of the three pillars - social, environmental, and economical - individually. The examples show that being conscious about sustainability can be guided by an agile approach. 

\subsubsection{The Social Pillar}
Often, we act as if our responsibility would end by delivering the value to the customer. An agile team typically aims to deliver regularly high value for the customer's advantage. After the (or rather after each) delivery, the team's  job is completed (except for maintenance and further development). However, if the team takes the full responsibility of their products, then they are also interested in the usage of the product and consider its impact to the world.

For example, some developers working for CHEF (a company providing a configuration management tool with the same name) were also interested in how their customers are using the product and how this is supporting the social good. In this case, these developers learned that one of their customers, the Customs and Border Protection or Immigration and Customs Enforcement (ICE), uses the product at the border between Mexico and the United States of America for running detention centres, ensuring deportation, and for implementing the family separation policy. In this case, the developers took the Brundtland definition (sustainability is about taking long-term responsibility for your action) and the social aim of ending poverty by heart and decided that the usage of their product does not confirm with their ethical values and has an unintended social impact. 

In the beginning, the developers brought their ethical interest to the attention of CHEF's management. However, the management at first referred to the long-standing contract and to the fact that the product has been used by ICE for many years (and nobody complained). The developers kept trying to convince the management that due to the change in politics also the usage of the product has shifted to the worse. However, the developers could only make a difference once one of the developers decided to delete all the code he contributed to the (Open Source) software. As a consequence, the product was not usable anymore for two weeks which created enough pressure for the management to decide on not renewing the contract \cite{7}.

It is important to understand that delivering value and satisfying the customer with your product is not an agile teams' sole responsibility. An agile team is also responsible for the social impact of the product it is creating. This means for a truly agile team, to stay in touch with the customer for recognising any differences in the usage of the product. This ongoing connection can be supported by automation such as monitoring, logging, and having tests that observe the usage of the product.

\subsubsection{The Environmental Pillar}
\label{sec:environmentalimpact}
The environmental pillar is mostly connected with the resources consumed. The global e-waste monitor reports that in 2016, 44.7 million metric tons of e-waste were generated – most often because the hardware gets (seemingly) too soon outdated \cite{8}. Additionally, as Nicola Jones reports, by 2030 information technology might exceed 21\% of the global energy consumption \cite{9}. 

Thus, it gets more and more important to consider the energy consumption of the products we are creating. Often it is assumed that hardware is cheap and thus, there is no need to pay much attention to performance, because if the software is not performing well enough we request that the hardware is getting faster. This proves Wirth's law from 1995 \cite{10}: ``Software is getting slower more rapidly than hardware becomes faster''. As elaborated by Gr\"{o}ger and Herterich, one of the reasons for this effect is the feature-creep where features are developed for a product that are unnecessary (not used) and don't fit the intended software architecture \cite{11}. 

The Mozilla foundation argues for the importance of examining cloud-based software in particular. In their 2018 Internet Health Report, they concluded that data centres have a similar carbon footprint as global air traffic with the latter being 2\% of all greenhouse gas emissions \cite{15}.  

However, while some companies ignore the problem despite the protests of their employees (see Amazon \cite{16}), others address it by shifting toward renewable energy for their data centres (see Google \cite{17}). This means for an agile organisation when deciding on a cloud infrastructure that it is essential to also consider the carbon footprint of that data centre. Therefore, it is the responsibility of an agile team to bring not only any technical information but also information about energy consumption and the carbon footprint of the infrastructure under question to the attention when the decision is up.

Another example is Mightybytes, a company focusing on developing digital strategies to create the design and user experience for their clients \cite{Frick2016}. By doing so, they pay particularly attention to how much energy is consumed by the designs they are creating and decide, for example, against including videos with high energy consumption. Additionally, they also take care of the environment the developers are in by ensuring the carpets are not toxic, there is enough space for everyone, natural light and plants, plus the offices are powered with renewable energy. Finally, other examples of initiatives that explore energy aspects in ICT can be found in \cite{Hilty2015b}, Part II.

\subsubsection{The Economical Pillar}
The economic dimension is often understood as the economic balance, e.g. that no nation (or company) grows economically at the cost of another one. Thus, topics like fair trade or paying fairly are often discussed along these lines. While this can be a topic also in (agile) software development, according to our experience this is seldom the case because, agile developers are still benefiting from good payments globally (however, this statement is not based on any research). Yet, there is another economic impact for organisations, because as the Cone Communications CSR study revealed, a company's reputation regarding their sustainability efforts will have an effect on both their market share and their search for talent \cite{18}. For example, as reported in this study:

\begin{quote}
    ``Nearly nine-in-10 Americans (89\%) would switch brands to one that is associated with a good cause, given similar price and quality, compared with 66 percent in 1993. And whenever possible, a majority (79\%) continue to seek out products that are socially or environmentally responsible''.
\end{quote}

An economic impact can also be made by organisations and teams through sharing learning. One example is Munich Re, one of the world's leading re-insurers, who got concerned about climate change already in the 70's. At that time, they began collecting and publishing research data about climate change. Protecting the research data for a competitive advantage was never considered by Munich Re because they realised that transparency allows them to learn from others and to improve the data. Transparency, they decided will increase both the general societal awareness of climate change and their own resilience \cite{19}. This insight provided a great foundation for Munich Re's further effort in combining also the other two pillars (environmental and social) with a general agile approach \cite{MunichRe2019, Jacobson}.

Transparency is also key for all the lessons learned in the near future on how to make the software we are creating more sustainable – only if we make those learning transparent right away, we can make a huge difference for everyone, everywhere.

\subsection{Company-wide Agility and Holistic Sustainability}
\label{sec:agilecompanies}
Implementing agility company-wide comes with a responsibility. Professionals as well as companies who claim to be agile are expected to also ``take actions based on the best interests of society, public safety, and the environment'' \cite{2}. Similarly, are corporate Agile Alliance members expected to ``to help make the software industry humane, productive, and sustainable'' \cite{1}. This means agile companies are expected to have a systemic view and understand the impact of the own actions and products created. This means, an organisation implementing company-wide agility has to have a wider perspective than one that is aiming at business agility only, as defined by the Business Agility Institute \cite{BAI}:
\begin{quote}
    ``Business agility is the capacity and willingness of an organisation to adapt to, create, and leverage change for their customer's benefit!''
\end{quote}

Thus, business agility focuses on the customer only whereas agile organisations aim for humanity and sustainability while having the society and the environment in mind. In this section we will explore companies that made quite some progress in implementing company-wide agility in that sense by having a holistic perspective on all three pillars, thus acting with social, environmental, and economical outcomes in mind.

\subsubsection{Patagonia}
Patagonia, Inc. is an American clothing company that markets and sells outdoor clothing since 1973. The company has become recognised as a leading industry innovator through its environmental and social initiatives, and the brand is now considered synonymous with conscious business and high-quality outdoor wear. In 2019, Patagonia received the 2019 Champions of the Earth award from the United Nations \cite{UN2019}, being recognised as an organisation that has sustainability at the very core of its successful business model. 

Patagonia's mission statement is ``We're in business to save our home planet''. They implement it by accomplishing a number of initiatives that inspires all levels of the organisation, as donating profits from their Black Friday sales (millions of dollars) to the environment through grassroots movements \cite{CNN2016}, or creating their new office space by restoring condemned building using recycled materials. The company states its benefits as: 1\% for the Planet; Build the Best Product with No Unnecessary Harm; Conduct Operations Causing No Unnecessary Harm; Sharing Best Practices with Other Companies; Transparency; and Providing a Supportive Work Environment. 

Patagonia has been cited as an example for Agile Organisations, not only because it has agile teams, but because they embody a north star across the organisation that recognises the abundance of opportunities and resources available, reducing the mindset of competition and scarcity and moving towards co-creating value with and for all of our stakeholders \cite{McKinsey2018}.

Other examples of their practices that demonstrate their concern about the three pillars are: encouraging consumers to think twice before making premature replacements, or over-consuming; designing durable textile yarns from recycled fabric; upholding a commitment to 100\% organic cotton sourced from over 100 regenerative small farms; sharing its best practices through the Sustainable Apparel Coalition's Higg Index; paying back an ``environmental tax'' to the earth by founding and supporting \textit{One Percent For The Planet}; donating its \$10 million federal tax cut to fund environmental organisations addressing the root causes of climate change. 

As an example of social impact, Patagonia invests on improving the supply chain to alleviate poverty. They screen their partners, as factories and more recently farms, using Patagonia staff, selected third-party auditors and NGO certifiers. They recognise a number of challenges, specially in the farm level \cite{Patagonia2016}, pp. 30: 
\begin{quotation}
``There can be land management and animal issues, as well as child labour, forced labour, pay irregularities, discrimination, and unsound health and safety conditions. These are often more difficult to resolve because of the complexities that extreme poverty, illiteracy and exploitation bring to this level of the supply chain. 

When it comes to land management, we're most concerned with a farm's use of chemicals and the impact its operations have on water, soil, biodiversity and carbon sequestration. For animal welfare, we look at humane treatment and slaughter. And when it comes to labour, we want to see safe and healthy working conditions, personal freedom, fair wages and honest payrolls.''
\end{quotation}

\begin{figure*}[h]
    \centering
    \includegraphics[width=1.00\textwidth]{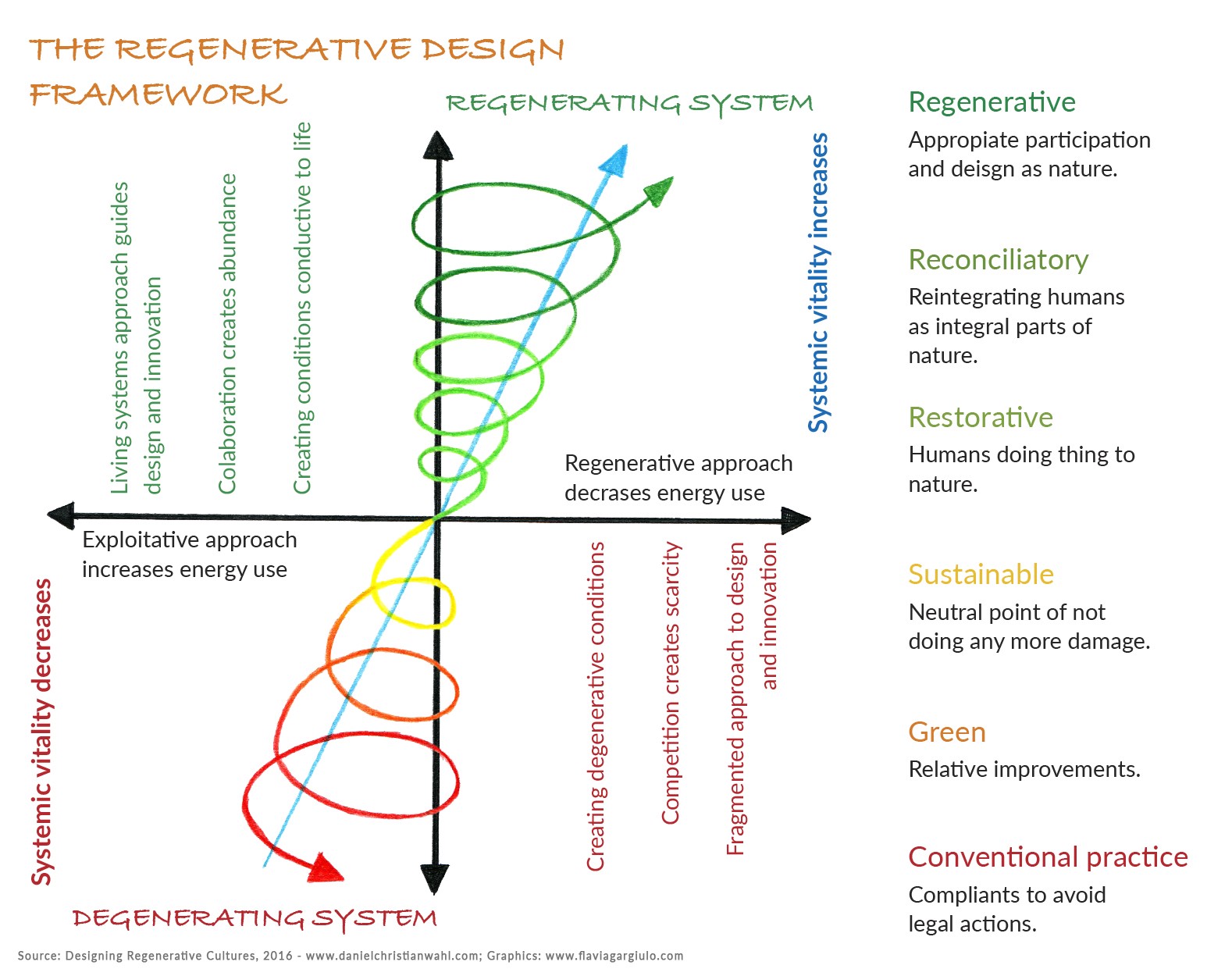}
    \caption{Continuum between Conventional, Sustainable and Regenerative practices \cite{Wahl2018}.}
    \label{fig:regenerative}
\end{figure*}

More recently, the company has started to support regenerative agriculture, establishing a goal of sourcing 100\% of their cotton and hemp from regenerative farming by 2030. It is important to stress how relevant this initiative is by introducing the meaning of regenerative. While the concept of \textit{sustainable} refers to a neutral point of not harming or damaging, the \textit{regenerative} concept goes beyond, stating that Humans are not only doing the right things to nature, but actually are an integral part of it, \textit{learning how to design as nature does} \cite{Wahl2018}. This means we can reverse the damage we've already done. Figure \ref{fig:regenerative} illustrates the continuum between 1) the conventional practices our society adopts in many areas, 2) sustainable practices, and 3) regenerative practices. 

Patagonia is a founding member of the Regenerative Organic Alliance \footnote{https://regenorganic.org/}, alongside Dr. Bronner's, Compassion in World Farming, Demeter and the Fair World Project and others. They created a certification that showcases whether a product has been made using processes to regenerate the land or not. 

Thus, supporting regenerative agriculture is a bold step Patagonia is taking that helps to reverse damage and create abundance. The reason is that agricultural practices are a huge contributor to climate change, accountable for around 25\% of global carbon emissions. Regenerative agriculture has the intention to restore highly degraded soil, enhancing the quality of water, vegetation and land-productivity altogether. It makes possible not only to increase the amount of soil organic carbon in existing soils, but to build new soil \cite{Rhodes2017}. If more companies follow this example, we will see more ecosystems being restored and communities being benefited.

\subsubsection{DSM-Niaga}
\begin{figure*}[h]
    \centering
    \includegraphics[width=0.9\textwidth]{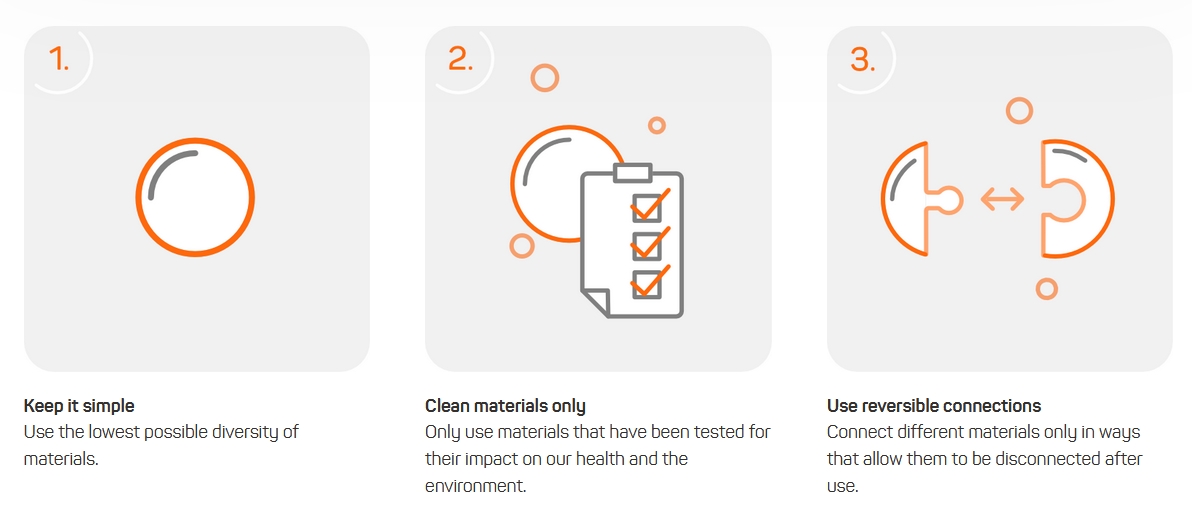}
    \caption{Design Principles of DSM-Niaga}
    \label{fig:designprinciples}
\end{figure*}

DSM-Niaga is a joint venture of the startup Niaga and the multinational firm Royal DSM. DSM-Niaga's vision is to design for circularity of everyday products. They started off with carpets and mattresses with the idea to stop these -most often toxic- products to go into land-fills but instead to decouple the material and use the very same material to go into the next production cycle. To guide this idea, they defined three design principles \cite{Niaga2020}, illustrated by Figure \ref{fig:designprinciples}:
\begin{enumerate}
    \item Keep it simple: Use the lowest possible diversity of materials.
    \item Clean materials only: Only use materials that have been tested for their impact on our health and the environment.
    \item Use reversible connections: Connect different materials only in ways that allow them to be disconnected after use. 
\end{enumerate}

For ensuring clean materials only and also proving it, they developed for example, a digital passport for every carpet based on blockchain technology. With the focus on the customer, this passport makes the complete value chain transparent. DSM-Niaga is also sharing their learning and pushing the industry to design for circularity:
\begin{quote}
    ``Moving forward, we will continue to focus our efforts and push boundaries to drive transparency and accountability across value chains. Indeed, with designers, producers and recyclers all needing to know what's in a product in order to recycle it, it's only a matter of time before digital product passports are in demand everywhere''.
\end{quote}


DSM-Niaga is focusing on both sustainability and company-wide agility. The firm is actually a teal organisation, that is a company defined by self-organisation where for example, employees are guided by the organisation's purpose and not by orders \cite{Laloux2014}. According to Rhea Ong Yiu, an Agile Coach at DSM-Niaga, the mother company (Royal DSM) is constantly learning from DSM-Niaga. Under the leadership of Feike Sijbesma, CEO and Chairman of the Managing Board, Royal DSM sold its entire petrochemical business. This has also been recognised. As a consequence, Feike Sijbesma has been appointed as Global Climate Leader for the World Bank Group, Co-Chair of the Carbon Pricing Leadership Coalition, and as Co-Chair of the Impact Committee of the World Economic Forum where he is one of the originators of the ``Stakeholder Principles in the COVID Era''. The latter states among other things \cite{WEF2020}:
\begin{quote}
    ``we must continue our sustainability efforts unabated, to bring our world closer to achieving shared goals, including the Paris climate agreement and the United Nations Sustainable Development Agenda''.
\end{quote}
Worth mentioning that the business strategy of Royal DSM (and as such as well of DSM-Niaga) is based on the United Nations Sustainable Development Goals. The sustainable development goals that are in particular focus for DSM-Niaga are \cite{Niaga2020b}:
\begin{itemize}
    \item Good health and well-being (Goal 3): Ensuring healthy lives and promoting well-being for all at all ages.
    \item Responsible consumption and production (Goal 12): Ensuring sustainable consumption and production patterns.
    \item Climate action (Goal 13): Taking urgent action to combat climate change and its impact.
\end{itemize}

\subsubsection{Sparda-Bank Munich}
Sparda-Bank is the largest cooperative bank in Bavaria, with more than 300,000 members. The bank maintains on its website a comprehensive description on how they implement all Economy for the Common Good (ECG) values, as well as their Common Good Balance Sheets and the certificate. Sparda was part of the first companies that agreed on ECG goals, back in 2010, being the first - and so far only - bank that operates according to the principles of the common good economy.

At the same time, the company keeps looking for innovation and agility \cite{Sparda2018}. So the organisation needs and it is opened to technological solutions. Due to its own values, it would have to carefully examine the need for and impact of the tools that it adopts. In fact, Sparda does have agile coaches and digitised solutions for their clients. They have also regularly had formats such as ``World-Café'' or smaller events with a ``marketplace character'' that are carried out in order to obtain the opinion of as many participants as possible and to initiate a dialogue. The design thinking method is adopted to support their product and projects solutioning. 

The company claims on its website \cite{Sparda2020b} that they are climate neutral and provides an annual CO2 balance. They reduce greenhouse gas emissions to the extent to what is technically and economically possible, or otherwise by purchasing climate certificates (or permits) in accordance with the Kyoto Protocol, which \cite{UNFCC}: 
\begin{quote}
    ``operationalizes the United Nations Framework Convention on Climate Change by committing industrialized countries and economies in transition to limit and reduce greenhouse gases (GHG) emissions in accordance with agreed individual targets [...] 
    
    One important element of the Kyoto Protocol was the establishment of flexible market mechanisms, which are based on the trade of emissions permits''. 
\end{quote}

Sparda-Bank has other initiatives that cover different aspects of the common good matrix. For instance, planting a tree for every new member or agreements to provide green electricity with special tariff for their clients. When financing an electric or hybrid car or an e-bike, Sparda-Bank Munich customers receive a reduced interest rate. They also state having no customer relationships with or investments in companies whose core business is in the armaments sector, as well as many other restrictions published on their website \cite{Sparda2020a}.

There are other examples related to suppliers and employees: they buy dishes and towels from works for the blind and disabled; they work to reduce the difference between the lowest salary and the highest salary (CEO)(which is currently published as 1:13.7 ratio); and finally they don't pay commissions or establish individual goals related to salary, only goals at a team level.


\section{Conclusion}
This paper examined that agility can contribute to sustainability. As we have seen, the Agile Manifesto in general and the principles in particular can provide guidance to sustainability \cite{12}. We have presented some achievements of the companies combining agility and sustainability. We explored examples of companies focusing on one of the dimensions only, as well as companies taking all three pillars into account. Certainly, if an agile company takes sustainability seriously, then it has to take an holistic view and look at all three dimensions at once - at people (social), planet (environmental), and profit (economic).

We have seen that most of these sample companies in the case studies are concentrating on using agile development and sustainability yet, without the focus on leveraging the one with the other. Especially the companies following a holistic approach implement sustainability by ICT. Using an agile approach for implementing sustainability in ICT seems to be a relatively new field.

\subsection{Criticism}
Sustainability became a trend and a symbol for progressive individuals, movements, and organisations, which sometimes leads to the so-called \textit{green-washing}. It happens when it seems to be important to have a reputation of being sustainable yet, not everyone who claims to live up to it really does it. Often this is supported by advertising for ones own sustainable reputation as \cite{Frick2016} exemplifies: 
\begin{quote}
    ``One hosting provider even claims in its marketing materials that it plants a tree for every new account, which is wonderful, but doesn't move us closer to an Internet powered by renewable energy''.
\end{quote}

Sustainability and its deeper implications are still not well-understood by society, despite more diffused, in particular because of the UN 2030 Agenda for Sustainable Development. Initiatives on many sectors, from business to NGOs and universities can easily distort or simplify it, intentionally or not. Taking the example of companies, we illustrated agile organisations that aim at balancing the 3 pillars in section \ref{sec:agilecompanies}, considering BCorps and ECG certified companies. It is important to be aware of the criticism (or limitations) around these models. 

The main critique to BCorps and similar movements is that these models still rely on capitalism as the core mechanism - and world view - for our economy, ``ignoring the possibility that capitalism itself, as it is largely practised today, might be at least one cause of the problems we are seeking to solve'' \cite{Coen2018}. This analysis is also supported by Nobel Prize winning economist Joseph Stiglitz \cite{Stiglitz2018}:
\begin{quote}
    ``Like the dieter who would rather do anything to lose weight than actually eat less, this business elite would save the world through social-impact investing, entrepreneurship, sustainable capitalism, philanthro-capitalism, artificial intelligence, market-driven solutions. They would fund a million of these buzzwordy programs rather than fundamentally question the rules of the game— or even alter their own behavior to reduce the harm of the existing distorted, inefficient and unfair rules''.
\end{quote}

High expectations are on digitalisation for sustainability by ICT. One example is that it is the replacement for paper, but, all digital products (thus, also the paper replacements) consume energy. Moreover, as Lorenz M. Hilty points out in \cite{11}:
\begin{quote}
    ``So far, neither in the case of air travel nor in the case of lifespan of household goods has the hope been fulfilled that due to digitalization the material and energy intensity of our activities would be reduced. It rather became evident in the digital age that providers turned the principle of intangible value creation through software into its opposite by stimulating or even forcing material consumption through software''.
\end{quote}
One reason is Wirth's law (Software is getting slower more rapidly than hardware becomes faster) that an update in software often requires the exchange of hardware \cite{10}. Another reason is the \textit{rebound effect}: the effect that environmental friendliness is a selling point that leads to overall higher consumption than before. And a third reason (related to the rebound effect) is, that the environmental friendly product is often used in addition - and not instead - to the environmental unfriendly product. An example for the latter are many car sharing offerings that are not used for substituting private cars but, instead, for substituting travelling by (local) public transport \cite{11}. 
Thus, the rebound effect is the main controversy for all achievements of digitalisation regarding sustainability.

\subsection{Outlook}
Although agile development promises sustainability for a long time, it has not been addressed sincerely thus far. However, there are some promising developments and also concrete ideas for delivering on agility's promise. Most importantly, we have to increase the awareness of the impact of agility so that at least agile teams can make conscious decisions on effecting the social, environmental, or economical dimension of sustainability.

\begin{figure}[h]
    \centering
    \includegraphics[width=0.5\textwidth]{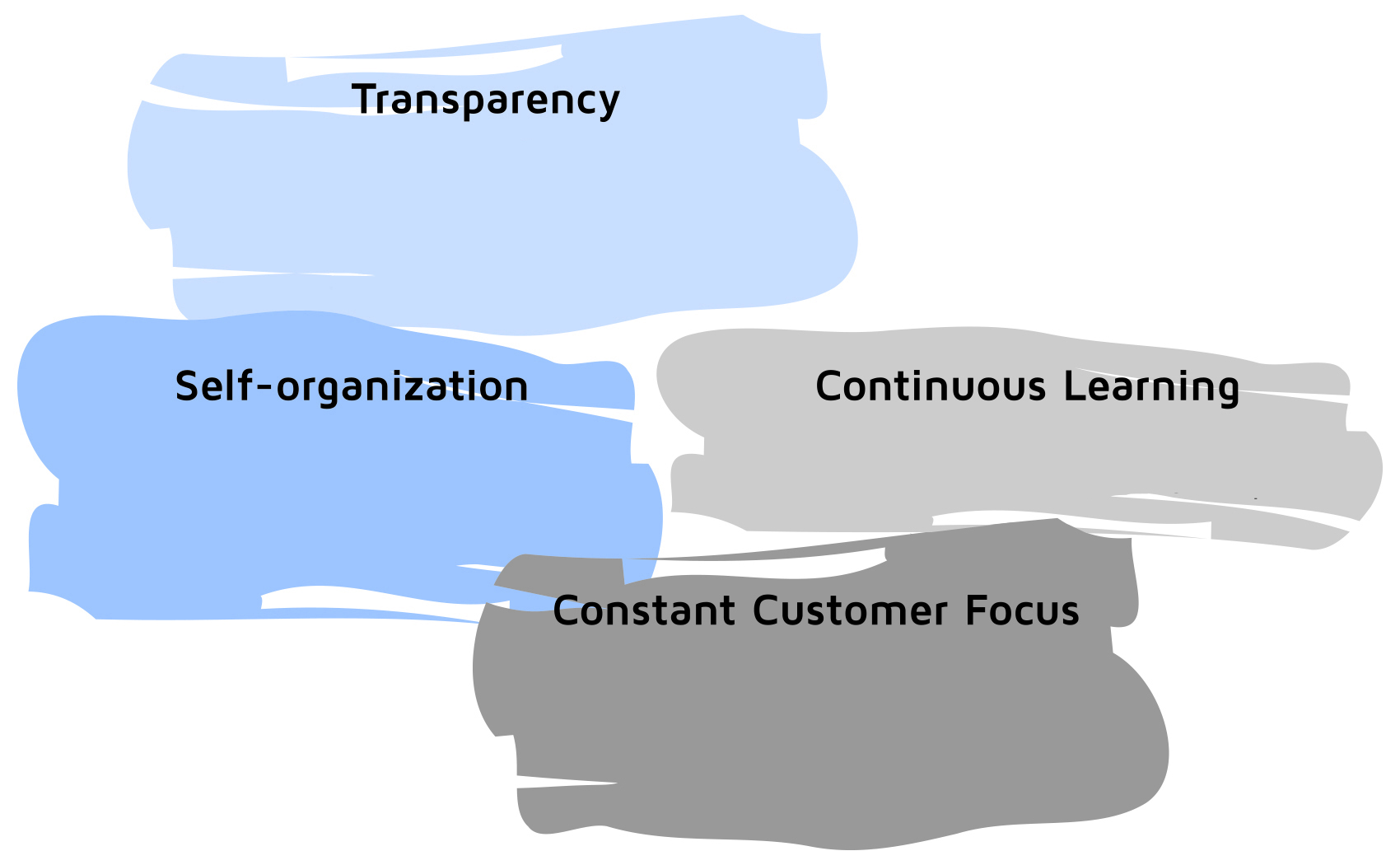}
    \caption{Four values guiding agile companies \cite{20}}
    \label{fig:4values}
\end{figure}

An agile team can for example, consider the energy consumption in their definition of done as well as via respective tests and monitoring. Individual teams and companies might discover ways to improve their own and their ecosystem's sustainability. Yet, only if these learning are shared and the effort improving sustainability is a collaborative one, can we really make a difference.
As pointed out by Eckstein \& Buck, a company claiming to be agile also has to aim for sustainability and as such needs to live up to the following values \cite{20} p. 196, also illustrated by Figure \ref{fig:4values}:
\begin{itemize}
    \item Self-organisation: An agile company should understand itself as a part of an ecosystem, belonging to itself, other companies, and the whole society.
    \item Transparency: An agile company makes its learning and doing transparent for the greater benefit for all. 
    \item Constant customer focus: An agile company understands all aspects of its ecosystem – be it social, environmental, or economic – as its customer.
    \item Continuous learning: An agile company learns continuously from and with its ecosystem to make the whole world a better place.
\end{itemize}

Fundamental for agile companies that are sustainability-aware is the need for a connected perspective \cite{20}, p. 198: ``This connected perspective incorporates the surrounding environment (economic, ecologic, societal, and social) in which companies operate.'' Thus, companies have to fulfil their role as active members of the society. One way for doing so, is by joining networks that focus on improving the economic, social, and environmental aspects of the society. Sample networks are: transparency international, global compact, fair labour association, or the climate group \cite{transparency, globalcompact, fairlabor, climategroup}.

Sustainability is not only important for (agile) companies because it's part of the Agile's vision \cite{1,2}. It is also important because it will be the key factor that decides on the survival of companies both in terms of finding talent and clients. This is the reason why some companies have already a sustainability officer in place who ensures sustainability in its many domains - environmental, economic, and social. It is the agile community's task to support the people in this role in making sustainability real.

With digitalisation becoming more momentum and the core competency of agile in software development more needs to be investigated in how agile development can make an important -positive- contribution for achieving higher sustainability. Because, as stated by the Karlskrona Manifesto \cite{Karlskrona2014}:
\begin{quote}
    ``Software in particular plays a central role in sustainability. It can push us towards growing consumption of resources, growing inequality in society, and lack of individual self- worth. But it can also create communities and enable thriving of individual freedom, democratic processes, and resource conservation''.
\end{quote}

\Urlmuskip=0mu plus 1mu\relax
\bibliographystyle{IEEEtran}
{
\bibliography{IEEEabrv,icse}}

\end{document}